\documentclass[aps,pra,twocolumn,groupedaddress]{revtex4-1}
\usepackage{times}
\usepackage{hyperref}
\usepackage[normalem]{ulem}
\usepackage{graphicx}
\usepackage{array}
\usepackage{amsmath}
\usepackage{amssymb}
\usepackage{color}
\usepackage{soul,xcolor}
\setstcolor{red}

\begin{document}
\title{Staggered quantum walks with superconducting microwave resonators}

\author{J. Khatibi Moqadam}
\affiliation{Instituto de F\'\i sica ``Gleb Wataghin'', Universidade Estadual de Campinas, Campinas, SP, Brazil}
\author{M. C. de Oliveira }\
\affiliation{Instituto de F\'\i sica ``Gleb Wataghin'', Universidade Estadual de Campinas, Campinas, SP, Brazil}
\author{R. Portugal}
\affiliation{Laborat\'orio Nacional de Computa\c c\~ao Cient\'\i fica (LNCC), Petr\'opolis, RJ, Brazil}

\date{\today}

\begin{abstract}
The staggered quantum walk model on a graph {is defined by an evolution operator that is the product of local
operators related to two or more independent graph tessellations. A graph tessellation is a partition of the set
of nodes that respects the neighborhood relation. Flip-flop coined quantum walks with the Hadamard or Grover coins
can be expressed as staggered quantum walks by converting the coin degree of freedom into extra nodes in the graph.
We propose an implementation of the staggered model with} superconducting microwave resonators, where the required
local operations are provided by the nearest neighbor interaction of the resonators coupled through superconducting
quantum interference devices. The tunability of the interactions makes this system an excellent toolbox for this
class of quantum walks. {We focus on the one-dimensional case and discuss its generalization to a more general
class known as triangle-free graphs.}
\end{abstract}


\maketitle

Quantum walks are the quantum generalization of random walks, and form the building blocks in designing quantum
search algorithms outperforming the similar classical ones~\cite{portugal2013quantum}. The two main paradigms
in this respect are the coined discrete-time quantum walk (DTQW)~\cite{ADZ93} and the continuous-time quantum
walk (CTQW)~\cite{FG98}.
In one-dimensional (1D) DTQWs, a two-level quantum system works as a coin, whose quantum property to exist in a
superposition of states gives the distinct {ballistic} spreading of the walker encoded in a set of
discrete states. In CTQWs, it is the excitation exchange between the neighboring sites, in a lattice, that directly
works as a walker without the need of a coin.
Typically a tight-biding Hamiltonian followed by a linear coupling between excitations in bosonic modes suffices to implement the CTQW model,
making {its} implementation  convenient (See Ref.~\cite{lozada2016quantum} for an example with nanomechanical resonators).
However, when the data structure is a lattice with dimension less than four, search algorithms based on the
standard CTQW do not outperform the classical algorithms based on random walks~\cite{childs2004spatial}.

{Recently, a general class of coinless discrete-time quantum walks was proposed---the staggered quantum
walk (SQW) ~\cite{portugal2016graph,Por16b,portugal2016staggered}, which includes the quantum walks studied
in Refs.~\cite{portugal2015one,Fal13} as particular cases. This model also includes as particular cases the flip-flop
coined DTQWs with Hadamard and Grover coins and the entire Szegedy's quantum walk model~\cite{Szegedy:2004}.}
In the language of graph theory, the required {unitary} operators {(not Hamiltonians)} can be obtained by a
graphical method based on graph
tessellations. {A tessellation is a partition of the set of nodes into cliques; that is, each element of the
partition is a clique. A clique is a subgraph that is complete, namely, all nodes of a clique are
neighbors.}

We have proposed an extension of the SQW model, called SQW with Hamiltonians~\cite{POM16}, which uses the
graph tessellations to define local Hamiltonians instead of the local unitary evolution operators.
The extended model includes the
quantum walks analyzed in Ref.~\cite{PRR05} as particular cases. The SQW with Hamiltonians is fitted for the implementation through bosonic nearest
neighbor interactions,
similarly to the CTQW, with the advantage of being able to outperform classical search algorithms at lower dimensional lattice structures~\cite{FP16}.
This
advantage comes at a price, which is the necessity to implement time dependent (piecewise-constant) controlled evolution,
requiring highly controllable systems for its implementation.

{Superconducting quantum circuits supporting microwave photons are promising for realizing the required
evolutions in quantum computation~\cite{Devoret2013superconducting,nori2016microwave,Barends2016} and quantum
simulation~\cite{houck2012chip,Georgescu2014quantum}.
Besides the tunneling devices employed for qubit encoding, superconducting circuits allow the realization of
lattices of coupled elements. Achieving tunable couplings between
circuit elements is a crucial step.
Tunable strong coupling among superconducting elements has been achieved in several ways, using both
superconducting quantum interference devices (SQUIDs) and qubits \cite{peropadre2013tunable,baust2015tunable,wulschner2015tunable,ploeg2007controllable,bialczak2011fast,chen2014qubit,geller2015tunable,hime2006solid,niskanen2006tunable,
niskanen2007quantum,yin2013catch,pierre2014storage,hoi2013giant,allman2014tunable,wu2016efficient}.
Recently, structured arrays of microwave superconducting resonators with SQUID tunable
couplings~\citep{peropadre2013tunable,baust2015tunable,wulschner2015tunable}
have been investigated on dedicated simulations of many-body models with engineered
interactions~\cite{benjamin2013bose,mei2013analog,deng2015sitewise,stassi2015quantum,deng2016Superconducting}.
The evolution of those systems cannot be simulated with conventional static-coupling quantum simulators.
Notwithstanding, those arrays could be employed for more general quantum tasks. Specially, an array of
microwave resonators with switchable couplings can be directly employed for simulating the SQW dynamics.


In this Letter, {we propose the implementation of the SQW model on a system composed of microwave
resonators coupled through SQUIDs. The {implementation} is analyzed {in details} on a 1D~lattice,
which {is used as a prototype to describe} a more general dynamics on triangle-free graphs.
In that class of graphs, which includes $N$-dimensional square lattices and trees, {the} resonators
interact {in a pairwise way} in each element of the tessellation.}
The conventional optical and electron-beam lithography technologies that are used in fabricating
superconducting-circuit-based devices allow to construct a large scale lattice with an arbitrary geometry. The
lattice dynamics can then be coherently controlled using external electromagnetic fields. Moreover, due to
typically large coherent times in superconducting circuits, more walk steps can be realized in such systems than
in any previous implementation~\cite{Manouchehri2014}.
\begin{figure}[t]
\includegraphics[trim = 0mm 60mm 0mm 0mm, clip=true, width=8.6cm]{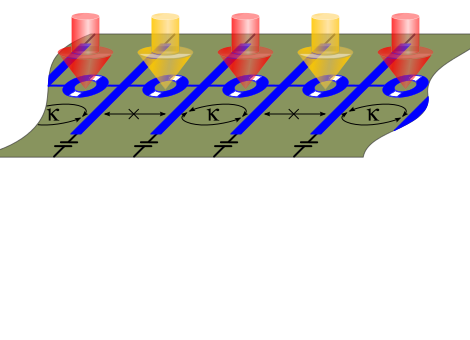}
\caption{\label{fig:system}(Color online) Array of superconducting microwave resonators coupled through
SQUID elements. Disjoint pairs of coupled resonators (with the coupling strength $\kappa$) are realized by applying
magnetic pulses with two different intensities, namely, the strong (red/dark-gray) and the weak (yellow/light-gray)
pulses on the SQUIDs. The SQW dynamics is implemented by alternating the application of pulses with a fixed period.}
\end{figure}

Let us consider a 1D array of coupled superconducting microwave resonators, as in Fig.~\ref{fig:system},
which can be made from finite sections of superconducting transmission line~\cite{houck2012chip} or
stripline~\cite{baust2015tunable,wulschner2015tunable}. The resonators couplings are mediated by SQUID elements.
Each SQUID is controlled by an individual wave generator that produces magnetic flux pulses providing the system
with tunable couplings \cite{NoteX}.
The Hamiltonian for the system is written as ($\hbar=1$)
\begin{equation}
\label{eq:tight-binding}
\mathcal{H} = \sum_{n} \omega_n  a^{\dagger}_n a_n -
              \sum_{\langle n,m\rangle} \kappa_{nm}(\mathrm{\Phi_\mathrm{ext}})
                ( a^{\dagger}_{n} a_{m} + a^{\dagger}_{m} a_n ),
\end{equation}
where $\omega_n$ are the resonators frequencies, $a^{\dagger}_n$ and $a_n$ are the creation and annihilation
operators satisfying $[a_n,a^{\dagger}_m]=\delta_{nm}$ and $\kappa_{nm}(\mathrm{\Phi_\mathrm{ext}})$ are the
flux dependent couplings between  adjacent ($m=n\pm1$) resonators. Hamiltonian (\ref{eq:tight-binding}) represents the tight-binding model with
controllable hopping strengths $\kappa_{nm}(\mathrm{\Phi_\mathrm{ext}})$.
Similar tunable coupling with SQUIDs has also been discussed in Refs.~\cite{peropadre2013tunable,deng2015sitewise}.

The SQUID coupler is specially appropriate, since it allows turning on and off the coupling between the two resonators,
hence working as a switch. Actually, two different states, namely, a given large coupling {$\kappa$} and
no coupling are required here. Such states can be implemented by applying two different pulses
through the corresponding wave generator, say {$\Phi_\mathrm{on}$ and $\Phi_\mathrm{off}$, such that
$\kappa_{nm}(\mathrm{\Phi_\mathrm{on}})=\kappa$ and ${\kappa_{nm}(\mathrm{\Phi_\mathrm{off}})=0}$}.
{Such {ability to switch on and off} the couplings is essential for our {model}.}

There are several methods to prepare the system in a {predefined} state of the resonators
and also to measure {their} state after the evolution. We are interested to describe a single-particle walker, therefore methods for single-photon generation and detection are
required ~\cite{johnson2010quantum,hofheinz2008generation,hofheinz2009synthesizing}.
In order to prepare and measure photons in an arbitrary resonator, individual transmon
qubits~\cite{koch2007charge} are coupled capacitively to the resonators.
Each transmon qubit is also coupled capacitively to a separate superconducting resonator---a coplanar waveguide
cavity, which is required for manipulating the qubit state.
The dynamics of the transmon qubit coupled to the $n$th resonator in near resonant regime is described by the
Jaynes-Cummings Hamiltonian. The way those additional devices are employed for photon generation and detection is described after exploring the system dynamics.

A SQW on the 1D lattice is defined by two tessellations described in Fig.~\ref{fig:lattice}~(a).
{The set of $N$ nodes of the array can be associated with the canonical
basis $\{|n\rangle:\,n=0\ldots N-1 \}$, where $|n\rangle$ is a $N$-component
unit vector with $1$ in the $(n+1)-$th entry and $0$ otherwise, spanning the $N$-dimensional Hilbert
space. We associate vectors $| \alpha_{n} \rangle = \bigl( | n \rangle + | n+1 \rangle \bigr) / \sqrt{2}$ with the 2-node elements (colored ovals in Fig.~\ref{fig:lattice}) of both tessellations and $| \alpha_{N-1} \rangle=  | N-1 \rangle$, $| \alpha_{N} \rangle=  | 0 \rangle$ with the 1-node elements.  Even (odd) $n$ refers to red (yellow) tessellation.
The Hamiltonian for the red (yellow) tessellation is
\begin{equation}
\label{eq:SQW_Hamiltonians}
\mathcal{H}_{0(1)} = 2 \sum_{\substack{ n\;\mathrm{even} \\ (n\;\mathrm{odd}) }}
|\alpha_{n} \rangle \langle \alpha_{n} | - {\mathcal{I}}_N,
\end{equation}
where $\mathcal{I}_N$ is the $N$-dimensional Hilbert space identity operator~\cite{POM16}.
The Hamiltonians are block diagonal, and each block is given by the Pauli matrix $\sigma_x$.
The local operator of SQW is defined as $\mathcal{U}_{0(1)} = \exp\left( i \theta \mathcal{H}_{0(1)} \right)$,
where $\theta$ is an  angle~\cite{POM16}. Since the Hamiltonians are block diagonal, the operators
are diagonal as well and the blocks are given by $\exp \left( i \theta \sigma_x \right)$.
The SQW dynamics is driven by successive applications of $\mathcal{U}_1\mathcal{U}_0$, starting from an initial state.
\begin{figure}[t]\includegraphics[trim = 0mm 0mm 0mm 0mm, clip=true, width=8.5cm]{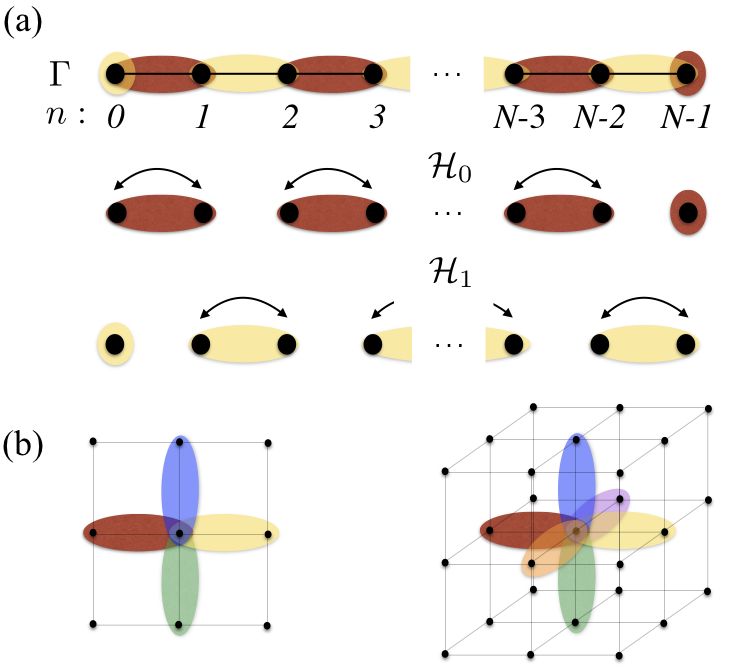}
\caption{\label{fig:lattice}(Color online) (a) A 1D array, with the two possible tessellations,
namely, the red/dark-gray and the yellow/light-gray ovals. Each tessellation is a partition of
the set of nodes into cliques (a clique is a subgraph in which every two distinct nodes are
connected by an edge). Moreover, the set of edges are covered in the union of all tessellations.
Each tessellation corresponds to a Hamiltonian. (b)~The ``unit cells'' for the
two- and three-dimensional lattices. For the $N$-dimensional lattice $2N$ different tessellations
are required. }
\end{figure}

The SQW dynamics can be achieved by controlling the superconducting circuit.
We consider Hamiltonian~(\ref{eq:tight-binding}) in the single-photon
regime $\sum_n \langle a^{\dagger}_n a_n \rangle = 1$. {Therefore the state of the system with $N$
resonators belongs to the $N$-dimensional Hilbert space, which is given in terms of the canonical
basis $\{|n\rangle\}$ previously described.} The resonators are considered  in resonance  {at} frequency~$\omega$.
{For the required} dynamics,  the Hamiltonians $\mathcal{H}_0$ and $\mathcal{H}_1$
in Eq.~(\ref{eq:SQW_Hamiltonians}) {are slightly modified} by substituting  $\sigma_x$  with 
$\omega \mathcal{I}_2 - \kappa \sigma_x$. 
The modified Hamiltonians {for} $N$  odd  \footnote{The description for even $N$ must be changed accordingly.} can be written in the explicit form
\begin{equation}
\label{eq:H0}
\mathcal{H}_0 = \begin{bmatrix}
\mathcal{I}_{(N-1)/2} \otimes (\omega \mathcal{I}_2 - \kappa \sigma_x) &{\bf{0}}\\
{\bf{0}} & \omega\\
\end{bmatrix},
\end{equation}
and
\begin{equation}
\label{eq:H1}
\mathcal{H}_1 = \begin{bmatrix}
\omega & {\bf{0}}\\
{\bf{0}} &
\;\;\;\mathcal{I}_{(N-1)/2} \otimes (\omega \mathcal{I}_2 - \kappa \sigma_x)\\
\end{bmatrix}.
\end{equation}
Non-commuting Hamiltonians $\mathcal{H}_0$ and $\mathcal{H}_1$ are referred to as even and odd,
respectively.
{Here} we explicitly consider open boundary conditions, and with a small modification periodic boundary conditions could also be addressed. Since we simulate the dynamics far from the boundaries,
corresponding to the walk on the line, this choice is not relevant.

The even (odd) Hamiltonian is constructed by switching on only the couplings $\kappa_{n,n+1}$ with even
(odd) index $n$, {directly corresponding to the tesselation} in Fig.~\ref{fig:lattice} {(a)}.
{We are} interested  {in controlling} the system by alternating between the even and the odd Hamiltonians,
in certain time steps $\tau$. Therefore, we apply the flux pulses
{$\Phi_\mathrm{on}$ and $\Phi_\mathrm{off}$ \cite{NoteX}}, such that in the time interval [0,2$\tau$) the couplings
assume the form
\begin{equation}
\label{eq:time_dependent_couplings}
 \begin{cases}
    \begin{array}{ll}
       \kappa_{n m}(\Phi_\mathrm{ext}) = \kappa, \\
       \kappa_{n p}(\Phi_\mathrm{ext}) = \kappa_{q m}(\Phi_\mathrm{ext}) = 0,\;\;\forall (p,q) \neq (m,n),
    \end{array}
 \end{cases}
\end{equation}
where $n=2\ell$ for $t \in [0,\tau)$, and $n=2\ell+1$ for $t \in [\tau,2\tau)$,
in which $\ell=0\ldots(N-1)/2$
(note that $m=n\pm 1$ for the 1D array).
{The system setup during $[0,\tau)$ is given in Fig.~\ref{fig:system}, where the red (yellow) magnetic pulses
associated with the flux $\Phi_\mathrm{on}$ ($\Phi_\mathrm{off}$). In $[\tau,2\tau)$, the magnetic pulses are
interchanged.} {The realization of} such time dependent couplings implies that the system is described by the
Hamiltonian
\begin{equation}
\label{eq:SQW}
\mathcal{H}(t) = \begin{cases}
                          \mathcal{H}_0, & t \in [0,\tau) \\
                          \mathcal{H}_1, & t \in [\tau,2\tau)
                         \end{cases},
\end{equation}
{which} generates the evolution operator
\begin{equation}
\label{eq:evolution}
\mathcal{U}(2\tau)=\mathcal{U}_1(\tau)\;\mathcal{U}_0(\tau),
\end{equation}
where~$\mathcal{U}_{0}$~($\mathcal{U}_{1}$) corresponds to {the evolution} of the time-independent
Hamiltonian~$\mathcal{H}_{0}$~($\mathcal{H}_{1}$).
$\mathcal{U}_{0}$ ($\mathcal{U}_{1}$) is {easily calculated from} the
exponential of the $2\times2$ matrices in the block diagonal Hamiltonian $\mathcal{H}_{0}$ ($\mathcal{H}_{1}$)
\begin{equation}
\label{eq:coin}
e^{ -i\tau ( \omega \mathcal{I}_2 - \kappa \sigma_x ) } = e^{-i \omega \tau}
\begin{bmatrix}
\cos\kappa\tau & i \sin\kappa\tau \\
i \sin\kappa\tau & \cos\kappa\tau \\
\end{bmatrix}.
\end{equation}
{The parameter $\theta$  introduced in the mathematical model is now set as
$\theta = \kappa \tau$,  by adjusting the time interval $\tau$, and as far as the resonators are in resonance, the role of $e^{-i \omega \tau}$ in (\ref{eq:coin}) is
irrelevant. 
This procedure
allows to implement a general 1D-SQW dynamics. For instance, by setting $\kappa\tau=2\pi \ell+\pi/4$,
for an integer $\ell$}, the blocks of the evolution operators~$\mathcal{U}_{0}$
and $\mathcal{U}_{1}$ take the form of the Hadamard-like operator
\begin{equation}
\label{eq:Hadamard}
H = \frac{1}{\sqrt{2}}
\begin{bmatrix}
1 & i  \\
i  & 1 \\
\end{bmatrix},
\end{equation}
{and the quantum walk model introduced in Ref.~\cite{PRR05} is recovered. To have the fastest spread
of the walker's probability distribution one must set  $\kappa\tau=2\pi \ell+\pi/3$~\cite{POM16}.

In the following, we consider the time evolution of the system for the period $t=2\tau l$ with $l$ an integer
number, under the repeated application of operator~(\ref{eq:evolution}), leading to $[\mathcal{U}(2\tau)]^l$.
The evolution starts with the initial state
\begin{equation}
\label{eq:initial_state}
|\psi_0\rangle = | (N-1)/2 \rangle,
\end{equation}
{representing} a single photon in resonator $(N-1)/2$, the middle resonator of the chain.
In order to produce such initial state, firstly all the couplings are turned off, and then a single photon
is generated in resonator $(N-1)/2$. 
Generating a single photon in a resonator of the system is possible by promoting the corresponding
transmon to its first excited state and then mapping the excitation into the resonator.
The protocol begins by exciting the transmon by applying a $\pi$-pulse through the coplanar waveguide (CPW) 
cavity,
while the qubit is detuned from the system resonator (CPW cavity and system resonator must have
different different frequencies). 
Then the transmon is brought to resonance with the system resonator for the
time $t_\mathrm{Rabi}=\pi/2\lambda$ {($\lambda$ is the qubit-resonator coupling strength)}
to transfer its excitation to the resonator.

Now the system evolves as
\begin{equation}
|\psi_l\rangle = [\mathcal{U}(2\tau)]^l | \psi_0 \rangle,
\end{equation}
for a given integer $l$.
At this stage, we measure the system by detecting all the resonators. That can be done by turning off
all the couplings, and then measuring the population of all the resonators.
A resonator photon number detection is processed by bringing the transmon  into resonance with the
resonator for $t_\mathrm{Rabi}=\pi/2\lambda${, hence,} the (excited) resonator transfer back the photon
to the qubit. Spectroscopy of the transmon transition frequency through far detuned CPW cavity then
gives the information about the photon number in the resonator.
Such measurement protocol, however, destroys the photon in the system resonator. To have a non-demolition
measurement of the photon number, the transmon should interact with the system resonator in a
quasi-dispersive regime~\cite{johnson2010quantum}.
In this case, the transition frequency of the transmon  is stark-shifted depending on the
number of photons, $0$ or $1$, in the system resonator. Now the spectroscopy of the transmon transition
frequency gives  information about the photon number in the system resonator,  in a non-demolition way.

Whatever the method employed, the probability distribution of finding the photon in the array is computed to give
\begin{equation}
P_l(n) = \Big| \langle n | \psi_l \rangle \Big|^2,
\end{equation}
for $n=0\ldots N-1$.
Figure~\ref{fig:prob_dist} shows the photon probability distribution for a linear array of $N=133$
resonators after $l=~32$ steps ($t=64\tau$), when  $\kappa\tau=2\pi l+\pi/3$ (maximum spread). The dynamics of the photon probability
distribution is ballistic---a clear signature of the quantum walk.
It should be mentioned that for obtaining the probability distribution in Fig.~\ref{fig:prob_dist} the above process
of initialization, evolution and measurement should be repeated many times. However, due to the
ballistic evolution of the quantum walk, the major part of the probability distribution is concentrated
around few resonators far from the initial position. Knowing that, the measurement stage can be
performed on a considerably smaller number of resonators.

\begin{figure}
\includegraphics[trim = 0mm 0mm 0mm 0mm, clip=true, width=8.6cm]{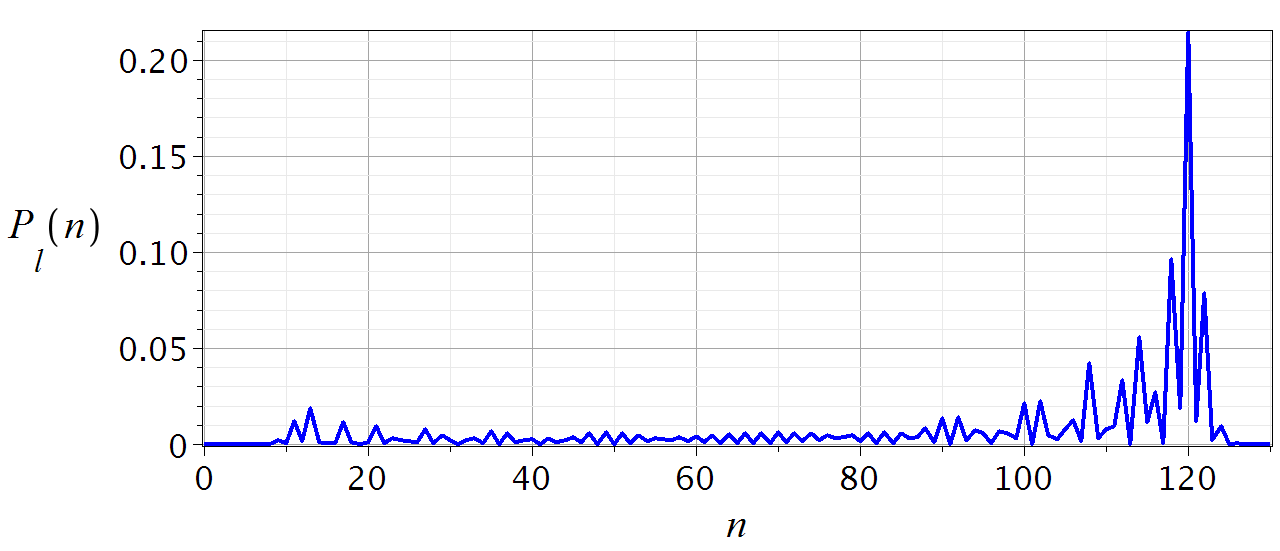}
\caption{\label{fig:prob_dist}(Color online) Photon probability distribution for a linear array of $N=133$
resonators after $l=32$ steps ($t=64\tau$). The photon is initially generated in the middle resonator of
the array and $\kappa\tau$ is tuned for maximum spread.}
\end{figure}

{
To conclude  we discuss the extension of the described 1D implementation to a class of graphs called
triangle-free graphs, which includes $N$-dimensional lattices, trees, and many other  topologies. A
graph is triangle-free if no three nodes form a triangle of edges.
To tessellate such a graph we make a partition of the node set by circling two neighboring nodes
at a time. Two different circles cannot have a node in common and no node can be missed at the end of the process
(there is the possibility of ending up with isolated single nodes that form singletons of the partition).
The red partition in Fig.~\ref{fig:lattice}~(a) is an example of this procedure. The circles are labeled then
by $\alpha_k$ for $0 \le k < c_0$, where $c_0$ is the number of circles in the partition.
Having related a Hilbert space basis to the node set, we associate the unit
vector $| \alpha_k \rangle = (|i\rangle+|j\rangle)/\sqrt 2$ with circle $\alpha_k$ that contains the nodes $i$
and $j$ (if $\alpha_k$ contains only node $i$ then $|\alpha_k\rangle=|i\rangle$). Accordingly, the Hamiltonian
\begin{equation}\label{eq:SQW_Hamiltonians_triangle_free}
  \mathcal{H}_0 = 2 \sum_{k=0}^{c_0-1} |\alpha_k\rangle\langle\alpha_k| - {\mathcal{I}}_N,
\end{equation}
is defined for the tessellation associated with $\{|\alpha_k\rangle\}$.
The vectors $|\alpha_k\rangle$ have at most two nonzero entries in the computational basis and
Hamiltonian $\mathcal{H}_0$ is a reflection operator~\cite{POM16}.
The same procedure is repeated to obtain a second tessellation, but the new partition must aim the edges that
were not included in the circles of the first partition. The new Hamiltonian $\mathcal{H}_1$ can be
obtained from Eq.~(\ref{eq:SQW_Hamiltonians_triangle_free}) after replacing $|\alpha_k\rangle$ with the vectors
associated with the second partition and replacing $c_0$ with $c_1$, where $c_1$ is the number of circles in the
second partition.
The process is continued until all edges have been covered with circles and the Hamiltonian $\mathcal{H}_{d-1}$
has been obtained. Besides, each node must be inside the intersection of $d$ circles. This situation can be seen
for dimensions higher than $1$ in the ``unit cells'' in Fig.~\ref{fig:lattice}~(b).

The evolution operator of a SQW with Hamiltonians in the class of triangle-free graphs has the form
\begin{equation}\label{eq:SQW_operator_triangle_free}
	\mathcal{U} = \textrm{e}^{i\theta \mathcal{H}_{d-1}}\,\cdots\,\textrm{e}^{i\theta \mathcal{H}_1}
	              \textrm{e}^{i\theta \mathcal{H}_0},
\end{equation}
where $\theta$ is an angle and $d$ is the maximum vertex degree, that is, the maximum number of edges incident on
a node. For $N$-dimensional lattices $d=2N$, that can be verified for $N=2,3$ in Fig.~\ref{fig:lattice}~(b).

According with our prescription any desired triangle-free graph can be implemented using resonators and SQUIDs associated with the nodes and the edges
of the graph, respectively. The system is described by Hamiltonian~(\ref{eq:tight-binding}), where the first sum
runs over all nodes and the second sum runs over all edges.
Each of the Hamiltonians $\mathcal{H}_0,\cdots,\mathcal{H}_{d-1}$ can be implemented during the time
period $\tau$ by applying an appropriate set of magnetic pulses, such that the couplings take the form of
Eq.~(\ref{eq:time_dependent_couplings}), in which $m,n$ belong to a suitable $\alpha_k$.
The corresponding setup of the system, in each case, consists of a collection of disjoint pairs of coupled resonators,
similar to the setup in Fig.~\ref{fig:system}. Therefore, the time-independent Hamiltonians can be implemented during
the time interval $[0,\tau d)$ leading to the evolution~(\ref{eq:SQW_operator_triangle_free}). 
We remark that quantum search algorithms~\cite{portugal2016graph} employing the present proposal can be readily implemented. For that an extra local Hamiltonian associated with a non-homogeneous tessellation is required.

Finally, considering that the coupling strength is about $10$ MHz \cite{NoteX} and the single photon lifetime
in the resonators is around $100$ $\mu$s or higher~\cite{ Devoret2013superconducting}, there is enough time to realize a
considerable number of steps. Moreover, the magnetic pulses should be switched within $0.1$ $\mu$s.
Imperfection in the resonators and couplings frequencies can affect the dynamics producing, for example, wavefunction
localization. However, it is expected that the system can tolerate small imperfections in the
couplings similarly to the continuous-time dynamics~\cite{lozada2016quantum}.

JKM acknowledges financial support from CNPq grant PDJ 165941/2014-6. MCO acknowledges support by FAPESP through the
Research Center in Optics and Photonics (CePOF) and by CNPq. RP acknowledges financial support from
Faperj (grant n.~E-26/102.350/2013) and CNPq (grants n.~303406/2015-1, 4741\-43/2013-9).


%

\clearpage
\onecolumngrid
\begin{center}
\textbf{Supplementary Material for ``staggered quantum walks with superconducting microwave resonators''}
\end{center}


\renewcommand{\theequation}{S\arabic{equation}}
\setcounter{equation}{0}
\renewcommand{\thefigure}{S\arabic{figure}}
\setcounter{figure}{0}

\section{Derivation of the system Hamiltonian}
Here, we derive formally Hamiltonian~(1) corresponding to the circuit in Fig.~1
in the main text, which follows by quantizing the classical Lagrangian of the circuit.

Consider a one-dimensional array of coupled superconducting microwave resonators, as in Fig.~\ref{fig:circuit}~(a).
The microwave resonator, here a transmission line resonator, can be considered as a two-wire line, each piece of
infinitesimal length of which can be modeled as a $LC$ circuit, with the inductance and capacitance per unit
length $l$ and $c$, respectively {as shown in Fig.~\ref{fig:circuit}~(b)}~\cite{pozar2005microwave}.
Considering the flux variable ${\psi}_n (x,t)$ along the transmission line resonator, the
corresponding Lagrangian is given by~\cite{girvin2011circuit}
\begin{equation}
\label{eq:lagrangian_resonator}
\mathcal{L}^{\mathrm{R}}_{n} = \int_{-L}^{L} \left[ \frac{c}{2} ( \partial_t {\psi}_n )^2 -
                                               \frac{1}{2l}      ( \partial_x {\psi}_n )^2  \right] dx,
\end{equation}
where $c$ and $l$ are considered position independent and, without loss of generality, supposed to be
identical for all the resonators.
\begin{figure}[h]
\includegraphics[trim = 5mm 0mm 15mm 0mm, clip=true, width=8.6cm]{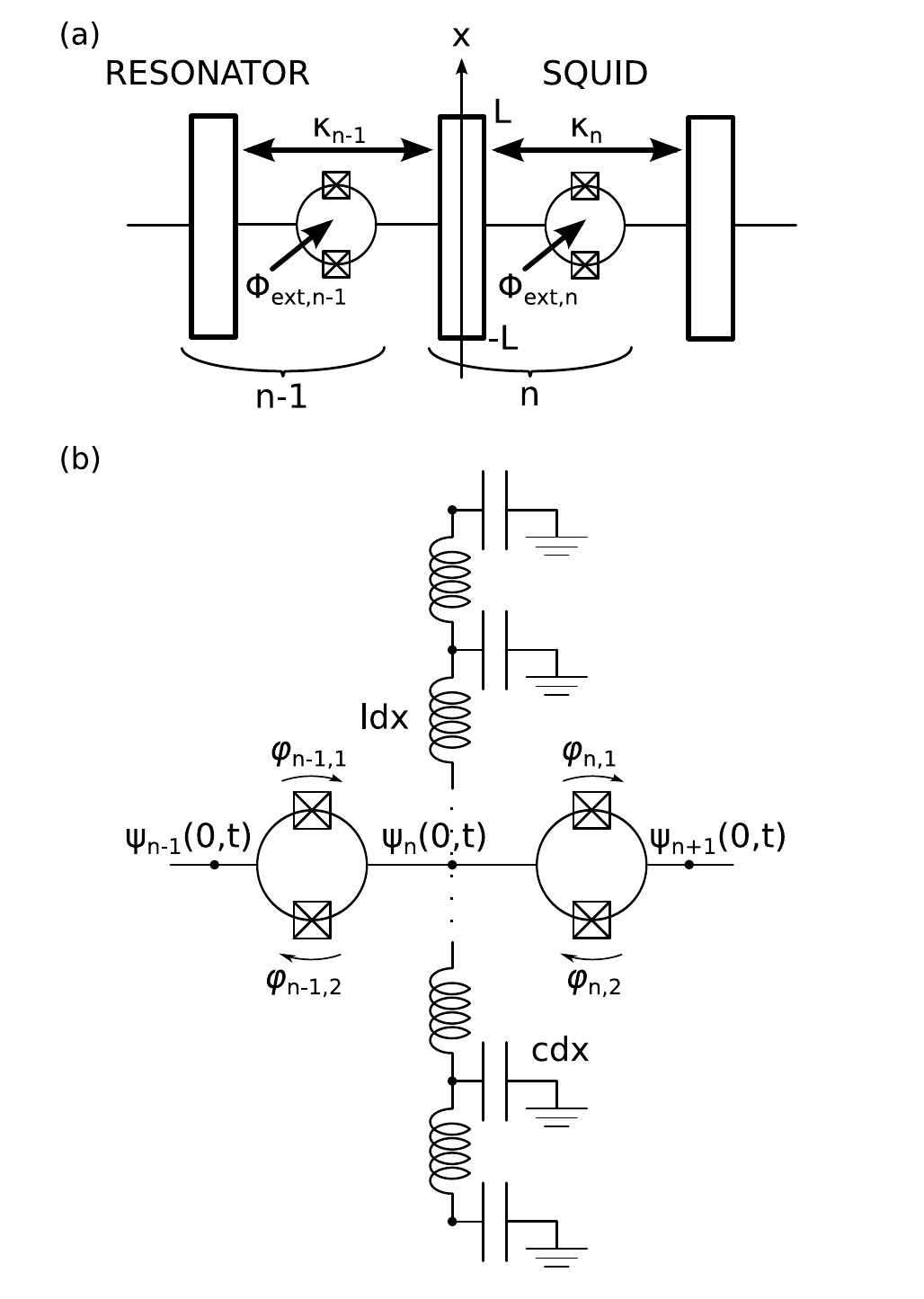}
\caption{\label{fig:circuit} Schematic representation of the system in panel (a) and the system parameters
with the lumped element model of the resonator in panel (b).}
\end{figure}
For a symmetrical SQUID, consisting of a superconducting ring interrupted by two identical Josephson junctions,
the Lagrangian is written as
\begin{equation}
\label{eq:lagrangian_SQUID}
\mathcal{L}^{\mathrm{S}}_{n} = \frac{1}{2}C_{J} ( \dot{\phi}^{2}_{n,1}+\dot{\phi}^{2}_{n,2} ) +
                                              E_{J} ( \cos \varphi_{n,1} + \cos \varphi_{n,2} ),
\end{equation}
where $C_{J}$ is the junction capacitance, $E_{J}$ is the Josephson energy and $\phi_{n,1(2)}$ and $\varphi_{n,1(2)}$
are the flux and the phase differences across the junctions, respectively.
The flux and the phase differences are related by $\varphi_{n,1(2)}=2\pi\phi_{n,1(2)}/\Phi_0$,
where $\Phi_0$ is the flux quantum. Again, without loss of generality, all the SQUIDs are assumed to have the
same $C_{J}$ and $E_{J}$.

The fluxoid quantization along the SQUID loop is given by $\phi_{n,1} + \phi_{n,2} + \Phi_{\mathrm{tot},n} = k \Phi_0$,
where $\Phi_{\mathrm{tot},n}$ is the total flux enclosed by the loop and $k$ is
an integer number~\cite{orlando1991foundations}. The total flux is the
sum of the externally applied flux $\Phi_{\mathrm{ext},n}$, subjecting the SQUIDs individually, and the flux generated by
the currents circulating through the loop. Here, it is assumed that the flux produced by the circulating currents is
negligible hence $\Phi_{\mathrm{tot},n}\approx\Phi_{\mathrm{ext},n}$. In this case, for the symmetrical SQUID,
it is possible to write $(\phi_{n,1} - \phi_{n,2})/2 = {\psi}_n (0,t)-{\psi}_{n+1} (0,t)$.
Now, the SQUID variables in Eq.~(\ref{eq:lagrangian_SQUID}) can be eliminated by expressing the Lagrangian 
in terms of $(\phi_{n,1} \pm \phi_{n,2})/2$.
Specifically, the second term in Eq.~(\ref{eq:lagrangian_SQUID}) changes to
$$2 E_J \cos \left( \pi \frac{\Phi_{\mathrm{ext},n}}{\Phi_0} \right) \cos \frac{2\pi}{\Phi_0}
   \left[ {{\psi}}_n(0,t)-{{\psi}}_{n+1}(0,t) \right],$$
supposing $k$ is an even integer. The flux-dependent cosine function can be expanded in terms of its
argument {$(2\pi/\Phi_0)[{\psi}_n(0,t)-{\psi}_{n+1}(0,t)]$ and only the first terms be kept,
when the argument is small. The small values for the argument correspond to small flux difference in the adjacent resonators}.
In the quantum regime, as discussed in the following,
the argument {is given in terms of the} creation and annihilation operators
{of the adjacent resonators} [see~Eqs.~(\ref{eq:general_solution})~and~(\ref{eq:coordinate_quantization})]
which is small for low photon numbers in the system. In this case, it is sufficient to consider the expansion up to
the second order terms {and dismiss} the higher order
terms {that correspond to nonlinear photon interactions}~\cite{peropadre2013tunable2}.
Therefore, the SQUID Lagrangian can be written as
\begin{align}
\label{eq:lagrangian_SQUID_modified}
\mathcal{L}^{\mathrm{S}}_{n} =
              C_{J} \left[ \partial_t{{\psi}}_n(0,t)-\partial_t{{\psi}}_{n+1}(0,t) \right]^2
              - E_n(\Phi_\mathrm{ext})
              \left[ {{\psi}}_n(0,t)-{{\psi}}_{n+1}(0,t) \right]^2,
\end{align}
where
\begin{equation}
\label{eq:E_n}
E_n(\Phi_\mathrm{ext}) =  \frac{4 \pi^2}{\Phi_0^2} E_J \cos \pi \frac{\Phi_{\mathrm{ext},n}}{\Phi_0},
\end{equation}
implying each SQUID can be controlled individually using the individual external magnetic fields.
Note that in Lagrangian~(\ref{eq:lagrangian_SQUID_modified}) all the terms independent of the flux variables
have been disregarded.

The system can be described by the Lagrangian
\begin{align}
\label{eq:lagrangian_system}
\mathcal{L} = \sum_n \tilde{\mathcal{L}}^{\mathrm{R}}_{n} + \mathcal{L}^{\mathrm{I}}_{n},
\end{align}
where, for each index $n$, the resonator Lagrangian $\mathcal{L}^{\mathrm{R}}_{n}$ and
SQUID Lagrangians $\mathcal{L}^{\mathrm{S}}_{n-1}$ and $\mathcal{L}^{\mathrm{S}}_{n}$ are considered.
The modified resonator Lagrangian is given by
\begin{align}
\label{eq:lagrangian_reonator_modified}
\tilde{\mathcal{L}}^{\mathrm{R}}_{n} = {\mathcal{L}}^{\mathrm{R}}_{n} +
 2 C_J [\partial_t {\psi}_{n}(0,t)]^2 - ( E_{n-1} + E_n ) [{{\psi}}_{n}(0,t)]^2,
\end{align}
that includes all the terms containing $\psi_n$, hence, the terms corresponding to
$\psi_{n\pm1}$ are left for $\tilde{\mathcal{L}}^{\mathrm{R}}_{n\pm1}$.
The interaction Lagrangian
\begin{align}
\label{eq:lagrangian_interaction}
\mathcal{L}^{\mathrm{I}}_{n} = -2 C_J \partial_t {\psi}_{n}(0,t) \partial_t {\psi}_{n+1}(0,t) +
                                2 E_n {\psi}_{n}(0,t) {\psi}_{n+1}(0,t)
\end{align}
includes the contributions that couple indices $n$ and $n+1$, hence, those that couple indices $n$ and $n-1$
are left for $\mathcal{L}^{\mathrm{I}}_{n-1}$.

Supposing the interaction {energy} between the {adjacent} resonators is
{small with respect to the energy of each resonator}, the problem can be treated
perturbativelly. In this way, the equation of motion is derived without considering the interaction Lagrangian
but then the corresponding solutions are used in the total Lagrangian that includes the interaction term.
The Euler-Lagrange equation for the modified Lagrangian $\tilde{\mathcal{L}}^{\mathrm{R}}_{n}$ turns out to be
\begin{align}
\label{eq:EL_equation}
\int_{-L}^L ( c \partial_t^2 {\psi}_n - l^{-1} \partial_x^2 {\psi}_n ) dx \; +
 4 C_J \partial_t^2 {\psi}_{n}(0,t) + 2 ( E_{n-1} + E_n ) {{\psi}}_{n}(0,t) = 0.
\end{align}

Away from the center of the resonator ($x=0$), Eq.~(\ref{eq:EL_equation}) gives the wave equation
$\partial_t^2 {\psi}_n = (1/\sqrt{lc})^2 \partial_x^2 {\psi}_n$, in which $1/\sqrt{lc}$ is the
velocity of the electromagnetic wave in the resonator.
By letting $\psi_n(x,t)=\xi_n(t) u_n(x)$ in the wave equation, the Sturm-Liouville equation
\begin{equation}
\label{eq:SL_equation}
\frac{d^2}{dx^2}u_n = - k_n^2 u_n,
\end{equation}
for the spatial mode is obtained. The equation corresponds to the $n$th resonator
in which $k_n = (\omega_n \sqrt{lc})$ is the wave number
and $\omega_n$ is the wave angular frequency.
At $x=0$, Eq.~(\ref{eq:EL_equation}) gives
\begin{align}
\label{eq:BC1}
 \left(\frac{du_n}{dx}\right)_{x=0^+} - \left(\frac{du_n}{dx}\right)_{x=0^-} =
 h_0 u_n(0),
\end{align}
which implies a discontinuity in the current passing through the resonator at $x=0$.
Moreover,
\begin{align}
h_0 =  2(-2 C_J \omega_n^2 + E_{n-1} + E_n)l
    = - 8 \chi_c k_n^2 L + \frac{{\chi_l}_{n-1}+{\chi_l}_n}{L}
\end{align}
in which the dimensionless parameters
\begin{equation}
\begin{cases}
\chi_c     &= C_J (2Lc)^{-1} \\
{\chi_l}_n &= E_n (2Ll)
\end{cases}
\end{equation}
are the relative capacitance and {the relative inverse} inductance of the Josephson junctions
with respect to the total resonator capacitance $2Lc$ and total resonator inductance $2Ll$.
Finally, at $x=\pm L$ we impose the open boundary condition (zero current)
\begin{align}
\label{eq:BC2}
 \left(\frac{du_n}{dx}\right)_{x=-L} = \left(\frac{du_n}{dx}\right)_{x=L} = 0.
\end{align}

The eigenfunctions of Eq.~(\ref{eq:SL_equation}) subjecting to the
constraint~(\ref{eq:BC1})~and~(\ref{eq:BC2}) can be written as
\begin{align}
\label{eq:spatial_eigenfunction}
u_{n,\nu}(x) = \begin{cases} 
                      A_{n,\nu} \cos k_{n,\nu}x - B_{n,\nu} \sin k_{n,\nu}x, -L \leq x \leq 0, \\
                      A_{n,\nu} \cos k_{n,\nu}x + B_{n,\nu} \sin k_{n,\nu}x, \;\;\;  0 \leq x \leq L,
               \end{cases}
\end{align}
where the wave numbers $k_{n,\nu}$ are the solutions of the transcendental equation
\begin{equation}
\label{eq:transcendental_eq}
\tan k_{n,\nu}L = \frac{h_0}{2k_{n,\nu}}=-4\chi_c k_{n,\nu}L + \frac{{\chi_l}_{n-1}+{\chi_l}_n}{2k_{n,\nu}L},
\end{equation}
and the integer $\nu$ labels different modes of oscillation.
Note that one of the coefficients in Eq.~(\ref{eq:spatial_eigenfunction}) is already known
in terms of the other one, say $B_{n,\nu}=h_0A_{n,\nu}/2k_{n,\nu}$.

The eigenfunctions~(\ref{eq:spatial_eigenfunction}), for a given resonator, form an orthogonal set for
different modes $\nu$ {according to}
\begin{align}
\label{eq:orthonormalization}
\frac{1}{2}c \int_{-L}^L u_{n,\nu}(x)u_{n,\nu^\prime}(x) dx +
2 C_J u_{n,\nu}(0)u_{n,\nu^\prime}(0) = \frac{1}{2}Lc\delta_{\nu,\nu^\prime},
\end{align}
which also determines the value of $A_{n,\nu}$.
Moreover, the derivatives of the eigenfunctions~(\ref{eq:spatial_eigenfunction}), obey the
{relation}
{
\begin{align}
\label{eq:orthonormalization_derivatives}
\frac{1}{2l}\int_0^L \frac{d}{dx}u_{n,\nu}\frac{d}{dx}u_{n,\nu^\prime}dx +
( E_{n-1} + E_n ) u_{n,\nu}(0)u_{n,\nu^\prime}(0) = \frac{1}{2}Lc\omega_{n,\nu}^2\delta_{\nu,\nu^\prime}.
\end{align}}

{
Figure~\ref{fig:eigenmodes} (a) shows the graphical solutions for equation~(\ref{eq:transcendental_eq})
for some typical values of $\chi_c$ and $\chi_l$. The first normal mode and its derivative which is
proportional to the current are sketched in Fig.~\ref{fig:eigenmodes} (b).
The frequency of the first normal mode corresponds to the smallest positive solution
of Eq.~(\ref{eq:transcendental_eq}). The discontinuity in the current is given by Eq.~(\ref{eq:BC1}) associating
with the current flowing to the adjacent resonator.}

\begin{figure}
\includegraphics[trim = 10mm 0mm 0mm 0mm, clip=true, width=8.5cm]{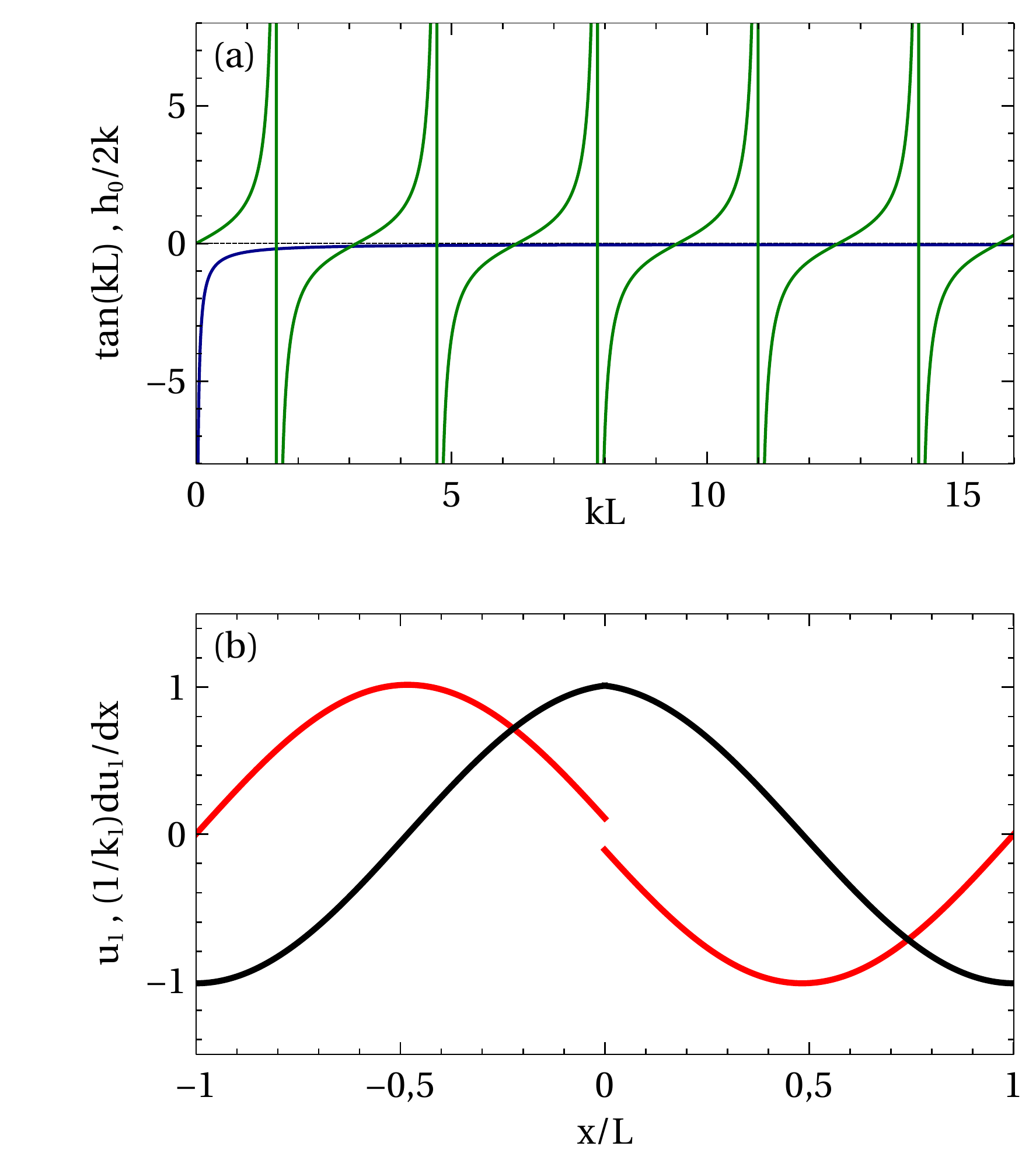}
\caption{\label{fig:eigenmodes}(Color online) Graphical solutions for equation~(\ref{eq:transcendental_eq})
with $\chi_c=0.5\times10^{-3}$ and ${\chi_l}_{n-1}={\chi_l}_{n}=-0.3059$ in panel (a), and the first normal mode and its
derivative (the current) in panel (b) (the plot with the discontinuity at $x=0$ corresponds to the current).
The frequency of the first normal mode is calculated as $k_{n,1}L = 3.0351$.}
\end{figure}

The general solution for Eq.~(\ref{eq:EL_equation}) is obtained by summing over
all normal modes $\xi_{n,\nu}(t) u_{n,\nu}(x)$, namely,
\begin{equation}
\label{eq:general_solution}
\psi_n(x,t)=\sum_{\nu} \xi_{n,\nu}(t) u_{n,\nu}(x),
\end{equation}
where each $\xi_{n,\nu}(t)$ is the temporal part of the wave function corresponding to the mode $\nu$.
Substituting the general solution~(\ref{eq:general_solution}) in the modified resonator Lagrangian
in Eq.~(\ref{eq:lagrangian_reonator_modified}), and using the the {relations}
(\ref{eq:orthonormalization}) and (\ref{eq:orthonormalization_derivatives}), we find
\begin{align}
\label{eq:lagrangian_harmonic_oscillator}
\tilde{\mathcal{L}}_n^{\mathrm{R}} = \frac{1}{2}Lc\sum_{\nu} \dot{\xi}_{n,\nu}^2 -
                                                             \omega_{n,\nu}^2 {\xi}_{n,\nu}^2,
\end{align}
which shows each (temporal) normal mode corresponds to an independent simple harmonic oscillator.
Considering the modes $\xi_{n,\nu}(t)$ as coordinates, the momentum conjugate to which are defined as
\begin{equation}
\label{eq:conjugate_momentum}
q_{n,\nu}(t) = \partial \tilde{\mathcal{L}}_n^{\mathrm{R}} / \partial \dot{\xi}_{{n},\nu}
= Lc \dot{\xi}_{{n},\nu},
\end{equation}
which can be used to write Hamiltonian~(\ref{eq:lagrangian_harmonic_oscillator}) as
\begin{equation}
\label{eq:hamiltonian_harmonic_oscillator}
\tilde{\mathcal{H}}_n^{\mathrm{R}} = \frac{1}{2}\sum_{\nu} \frac{1}{Lc} q_{n,\nu}^2 +
                                                            Lc\omega_{n,\nu}^2 {\xi}_{n,\nu}^2.
\end{equation}

Moreover, substituting the general solution~(\ref{eq:general_solution}) in the interaction
Lagrangian~(\ref{eq:lagrangian_interaction}) and using the conjugate
momentums~(\ref{eq:conjugate_momentum}) gives
\begin{align}
\label{eq:hamiltonian_interaction}
{\mathcal{H}}_n^{\mathrm{I}} = \sum_{\nu} g_{n}^\mathrm{cap}& q_{{n},\nu} q_{n+1,\nu}
                                         +g_{n}^\mathrm{ind}{\xi}_{{n},\nu}{\xi}_{n+1,\nu},
\end{align}
where
\begin{align}
g_{n}^\mathrm{cap} &= -\frac{2 C_J}{(Lc)^2} u_{n,\nu}(0) u_{n+1,\nu}(0), \\
g_{n}^\mathrm{ind} &= 2 E_{n} u_{n,\nu}(0) u_{n+1,\nu}(0),
\end{align}
and we have neglected the terms {that couple any pair of different modes
in the adjacent} resonators.

Hamiltonians~(\ref{eq:hamiltonian_harmonic_oscillator}) and (\ref{eq:hamiltonian_interaction})
can be quantized by introducing the creation and annihilation
operators, $a^{\dagger}_{n,\nu}$ and $a_{n,\nu}$, respectively, corresponding to the excitations
in mode $\nu$ in resonator $n$. The operators obey the commutation
relations $[a_{n,\nu},a^{\dagger}_{m,\nu^\prime}]=\delta_{nm}\delta_{\nu,\nu^\prime}$.
The coordinates $\xi_{{n},\nu}$ and the momentums ${q}_{{n},\nu}$ are then expressed
as~{\cite{girvin2011circuit}}
\begin{align}
\label{eq:coordinate_quantization}
\hat{\xi}_{{n},\nu} &= \sqrt{\frac{\hbar}{2Lc\omega_{{n},\nu}}} ( a_{n,\nu} + a^{\dagger}_{n,\nu} ), \\
\hat{q}_{{n},\nu} &= -i\sqrt{\frac{\hbar Lc \omega_{n,\nu}}{2}} ( a_{n,\nu} - a^{\dagger}_{n,\nu} ),
\end{align}
that turn Eq.~(\ref{eq:hamiltonian_harmonic_oscillator}) into
\begin{equation}
\label{eq:hamiltonian_harmonic_oscillator_multimode}
\tilde{\mathcal{H}}_n^{\mathrm{R}} = \sum_{\nu} \hbar \omega_{n,\nu}
          ( a^{\dagger}_{n,\nu}a_{n,\nu} + \frac{1}{2}),
\end{equation}
which is the quantized harmonic oscillator Hamiltonian with infinite non-interacting modes.
However, we restrict the harmonic oscillator to the first frequency mode,
hence, the subindex $\nu$ and the corresponding summation is dismissed and the resonator Hamiltonian becomes
$\tilde{\mathcal{H}}_n^{\mathrm{R}} = \hbar \omega_n a^{\dagger}_{n}a_{n}$,
in which the zero-point energy is also dropped to simplify.

The interaction Hamiltonian~(\ref{eq:hamiltonian_interaction}) is also quantized as
\begin{align}
{\mathcal{H}}_n^{\mathrm{I}} = \hbar\kappa_n^\mathrm{cap} (a_n - a^\dagger_n)(a_{n+1} - a^\dagger_{n+1}) +
            \hbar\kappa_n^\mathrm{ind} (a_n + a^\dagger_n)(a_{n+1} + a^\dagger_{n+1}),
\end{align}
where
\begin{align}
\label{eq:kappa_cap}
\kappa_n^\mathrm{cap} &= 2\chi_c u_{n}(0) u_{n+1}(0) \sqrt{\omega_n \omega_{n+1}}, \\
\label{eq:kappa_ind}
\kappa_n^\mathrm{ind} &=   \frac{{\chi_l}_n}{2k_{n}k_{n+1}L^2} u_{n}(0) u_{n+1}(0) \sqrt{\omega_n \omega_{n+1}},
\end{align}
and we have considered just the first mode of the resonators.
However, as mentioned before, the coupling between the resonators is weak. Moreover, the resonators
are assumed to be similar and in resonance. So, we can make the rotating wave approximation (RWA) discarding
the ``counter rotating terms'', $a_n a_{n+1}$ and $a^\dagger_n a^\dagger_{n+1}$, to obtain
\begin{equation}
{\mathcal{H}}_n^{\mathrm{I}} = -\hbar\kappa_n(\mathrm{\Phi_\mathrm{ext}}) ( a_n a^\dagger_{n+1} + a^\dagger_{n+1} a_n ),
\end{equation}
where
\begin{equation}
\label{eq:kappa_total}
\kappa_n(\mathrm{\Phi_\mathrm{ext}}) = -\kappa_n^\mathrm{ind}(\mathrm{\Phi_\mathrm{ext}}) + \kappa_n^\mathrm{cap},
\end{equation}
and we have stressed the external field dependency of the couplings by including it in the corresponding arguments.
The total Hamiltonian of the system is therefore appeared as in Eq.~(1) in the main text, after replacing
$\kappa_n$ with $\kappa_{nm}$ where $m=n \pm 1$ [see Fig.~\ref{fig:circuit}~(a)].

{
\section{Numerical data for the system frequencies}
For a transmission line resonator, an
impedance~$Z=\sqrt{l/c} = 50$~$\Omega$ \cite{peropadre2013tunable2,wulschner2015tunable2,goppl2008coplanar}
may be corresponded to a capacitance per unit length~$c=10^{-10}$ F/m~\cite{goppl2008coplanar} and the
impedance per unit length~$l=2.5\times 10^{-7}$~H/m. Considering the SQUID junctions
capacitance $C_J = 10^{-15}$~F~\cite{koch2010time-reversal}, for a resonator of
length $L=10^{-2}$~m (comparable with the microwave wavelength), we obtain $\chi_c = 0.5 \times 10^{-3}$.
On the other hand, the maximum value of $E_n(\Phi_\mathrm{ext})$ in Eq.~(\ref{eq:E_n}) is given
by $(4 \pi^2 / \Phi_0^2) E_J$ which can be calculated using $E_J=6.6262 \times 10^{-24}$~J for the Josephson
energy~\cite{koch2010time-reversal} and $\Phi_0=2.0679\times10^{-15}$~Wb for the flux quantum. Therefore
the maximum value of ${\chi_l}_n$ is obtained as
$\left|{\chi_l}_n\right|_\mathrm{max} = (4 \pi^2 / \Phi_0^2) E_J (2Ll) = 0.3059$.

Actually, the above values for $\chi_c$ and ${\chi_l}_n$ have been used in generating the plots
in~Fig.~\ref{fig:eigenmodes}, which give $k_{n,1}L = 3.0351$ corresponding to the first mode frequency.
Supposing all the resonators are identical, or in resonance with the same frequency $\omega$, we
obtain $\omega=607.028$~MHz.
The capacitive and inductive couplings in Eqs.~(\ref{eq:kappa_cap})~and~(\ref{eq:kappa_ind}) are then obtained as 
$\kappa^\mathrm{cap}=0.6193$~MHz and $\kappa^\mathrm{ind}=-10.2821$~MHz, respectively, for $u(0)= A = 1.01$.
}

{
\section{Switching on and off the couplings}
The time-dependent couplings required for the one-dimensional staggered quantum walk are given
in Eq.~(5). Such couplings lead to a collection of disjoint pairs of coupled
resonators at each time interval~$\tau$.
We can set $\kappa$ in Eq.~(5) to be the maximum value
of $\kappa_n(\mathrm{\Phi_\mathrm{ext}})$ in Eq.~(\ref{eq:kappa_total}), by setting
${\chi_l}_n = -\left|{\chi_l}_n\right|_\mathrm{max}$, as calculated in the previous section.
That corresponds to applying an external magnetic flux equal to the quantum flux
\begin{equation}
\kappa = \kappa_n(\mathrm{\Phi_\mathrm{ext}})\left|_{\Phi_{\mathrm{ext},n}=\Phi_0} \right.,
\end{equation}
hence $\Phi_{\mathrm{on}}=\Phi_0$.

The required external fluxes for turning off the
couplings $\kappa_{n\pm1}(\mathrm{\Phi_\mathrm{ext}})$, as demanded by Eq.~(5),
and given by Eq.~(\ref{eq:kappa_total}), are calculated by letting ${\chi_l}_{n\pm1} = 4 \chi_c (k_{n}L)^2$
[see Eqs.~(\ref{eq:kappa_cap})~and~(\ref{eq:kappa_ind})], where
we have assumed that all the resonators are identical. Using such value for ${\chi_l}_{n\pm1}$ modifies the resonator
frequency $\omega$ that was calculated in the previous section for the case all the couplings were on.
To obtain the new frequency for the case corresponding to a collection of disjoint pairs of coupled resonators,
the values for ${\chi_l}_{n-1}$ and ${\chi_l}_{n}$ should be substituted in the
transcendental Eq.~(\ref{eq:transcendental_eq}).
Having considered the first mode frequency, we get the system frequencies as $\omega=617.8077$ MHz,
$\kappa_n^\mathrm{ind} = -10.0054$ MHz, $\kappa_{n}^\mathrm{cap} = \kappa_{n\pm1}^\mathrm{cap} = 0.6242$ MHz and
$\kappa_{n\pm1}^\mathrm{ind}=\kappa_{n\pm1}^\mathrm{cap}$. Moreover, in this case, $k_{n}L=3.089$ which leads
to ${\chi_l}_{n\pm1} = 0.0191$, therefore, $\cos \left( \pi \Phi_{\mathrm{ext},n}/\Phi_0 \right)=0.0624$ and we obtain
\begin{equation}
0 = \kappa_{n\pm1}(\mathrm{\Phi_\mathrm{ext}})\left|_{\Phi_{\mathrm{ext},n}=0.4801\Phi_0} \right.,
\end{equation}
hence $\Phi_{\mathrm{off}}=0.4801\Phi_0$.}

{When the system topology corresponds to a general triangle-free graph, with degree $d$, each resonator
is coupled to $d$ resonators through $d$ SQUIDs. Therefore, the whole derivation for the 1D array (a triangle-free
graph with $d=2$) is valid here, but, slightly modified to include the extra SQUIDs coupled to each resonator. Finally,
an isolated pair of coupled resonators can be realized by turning on an specific coupling and turning off the remaining
$d-1$ ones.}

\end{document}